\documentclass[hidelinks]{article}
\usepackage[a4paper, total={5.5in, 8.5in}]{geometry}
\usepackage{appendix}
\usepackage{amssymb}
\usepackage{booktabs}
\usepackage{hyperref}
\usepackage{graphicx} 
\usepackage{amsthm}
\usepackage{amsmath}
\usepackage{amsfonts}
\usepackage[authoryear]{natbib}
\usepackage{bbm}
\usepackage{enumitem}

\usepackage{caption}
\usepackage{subcaption}

\usepackage{tcolorbox}
\usepackage{authblk}
\usepackage{wrapfig}
\newtcolorbox{mybox}{colback=blue!5!white,colframe=blue!75!black}
\usepackage{orcidlink}
\usepackage{hyperref}
\providecommand{\keywords}[1]
{
  \small	
  \textbf{\textit{Keywords---}} #1
}
\title{The Sequential Nature of Science: \\
Quantifying Learning from a Sequence of Studies}

\author[a]{Jonas M. Mikhaeil \orcidlink{0000-0001-6745-7505}\thanks{Address correspondents via E-mail to j.mikhaeil@columbia.edu}}
\author[b]{Donald P. Green}
\author[a]{David Blei}

\affil[a]{Department of Statistics, Columbia University, 1255 Amsterdam Ave, New York, NY 10027}
\affil[b]{Department of Political Science, Columbia University, 420 W. 118th Street, New York, NY 10027}

\begin{document}

\maketitle

\begin{abstract}
Scientific progress is inherently sequential: collective knowledge is updated as new studies enter the literature. We propose the sequential meta-analysis research trace (SMART), which quantifies the influence of each study at the time it enters the literature. In contrast to classical meta-analysis, our method can capture how new studies may cast doubt on previously held beliefs, increasing collective uncertainty. For example, a new study may present a methodological critique of prior work and propose a superior method. Even small studies, which may not materially affect a retrospective meta-analysis, can be influential at the time they appeared. To contrast SMART with classical meta-analysis, we re-analyze two meta-analysis datasets, from psychology and labor economics. One assembles studies using a single methodology; the other contains studies that predate or follow an important methodological innovation. Our formalization of sequential learning highlights the importance of methodological innovation that might otherwise be overlooked by classical meta-analysis.
\end{abstract}
\section{Introduction}
Scientific meta-analysis aggregates study results to summarize the current state of knowledge, and it helps us understand the degree of consensus in a scientific literature. However, classical meta-analysis does not reveal how we arrived at this consensus, nor does it identify which studies were particularly influential at the moment that they were published.

Intuitively, a study may influence consensus in different ways.  It might reduce uncertainty by presenting robust evidence from a large-scale study using an established methodology. Or it might cast doubt on previous estimates by introducing a new methodology or by reporting surprising results.

In this paper, we posit that answering these questions requires a sequential approach to meta-analysis, one where we consider the order in which the studies appeared to track how each one affected our understanding. Because we are tracking \textit{beliefs}---in the sense that a meta-analysis estimate characterizes the collective beliefs of a scientific community---we need to account for our uncertainty and how it changes over time.

To these ends, we propose the \textit{sequential meta-analysis research trace} (SMART). SMART takes in a sequence of studies that target the same unknown quantity. With these inputs, it produces a sequential Bayesian meta-analysis of the estimate and quantifies the contribution of each study to the state of knowledge. SMART aligns with the end results of classical meta-analysis, but also provides insight into the path taken to reach the current consensus.

SMART begins with an assumed Bayesian meta-analysis model, such as the classical random-effects meta-analysis model \citep{Higgins_Thompson_Spiegelhalter_2009}. Under the chosen model, it considers each study sequentially, quantifying each one's contribution by measuring its effect on the posterior. Specifically, we use the Wasserstein distance \citep{Villani_2008} to measure the change, which captures movement in both the mean estimate and the uncertainty surrounding it. SMART thus gives a sequential picture of how a series of studies have led to the current state of knowledge, and the relative contribution of each to that picture.

When evaluating a sequence of scientific studies, the methodology of a study can be as important as its specific empirical results. 
 To this end, we extend the classical random effects model to one that also accounts for which studies used the same methodology, e.g., propensity score matching or randomized trials. Our \textit{labeled random effects model} assumes that each methodology comes with an unknown bias, with a belief that more reliable methodologies have smaller biases. When incorporated into SMART, this new model allows for the possibility that studies using a superior methodology can make a larger contribution to the consensus estimate. In particular, it both informs our estimate of the biases of \textit{other} methodologies, as well as providing new estimates of the target itself.

 To demonstrate SMART, consider \citep{Card_Krueger_94}, Card and Krueger's seminal 1994 study on the effect of a minimum wage on employment. This study is part of a larger research literature that seeks to estimate this effect, and it was particularly important in its use of difference-in-differences together with a border discontinuity design to avoid confounding by changes in employment prospects.  Specifically, we focus on the beginning of the New Minimum Wage Research literature \citep{Ehrenberg_92}, a research program aimed at leveraging natural experiments to provide more credible estimates of the effect of the minimum wage on employment. Figure
\ref{fig:MW_figure} (Left) shows the estimates of this literature, as characterized by \citep{Neumark_Shirley_22}, leading up to Card and Krueger's study.
Using the labeled random effects model, Figure \ref{fig:MW_figure} (Right) shows the result of SMART, this literature's research trace. This trace tracks the sequential research contribution -- measured in terms of our proposed learning metric -- for every study as it enters the literature. 
While classical meta-analysis accords Card and Krueger's study little weight from a retrospective vantage point, SMART suggests that their study was highly influential.

Our approach contributes to the literature on sequential meta-analysis~\citep{Higgins_Whitehead_Simmonds_2011,Van_Der_Tweel_Bollen_2010,Spence_Steinsaltz_Fanshawe_2016,Bollen_Uiterwaal_Van_Vught_Van_Der_Tweel_2006,Kulinskaya_21}.  This literature has focused on methods that guarantee valid sequential inference \citep{Higgins_Whitehead_Simmonds_2011} or try to assess whether enough evidence has been collected for the results to be conclusive~\citep{Spence_Steinsaltz_Fanshawe_2016,Bollen_Uiterwaal_Van_Vught_Van_Der_Tweel_2006}. Recent work \citep{Kulinskaya_21} has analyzed the properties of sequential meta-analysis in the presence of temporal trends. 
In contrast, we provide a tool that allows us to track scientific consensus as it evolves.
Further, we extend the well-established random-effects model for meta-analysis \citep{Higgins_Thompson_Spiegelhalter_2009,Higgins_Whitehead_Simmonds_2011} to incorporate information and form inferences about the methodologies' biases.
This is similar to the mixed linear model developed in \citep{Raudenbush_1985} which incorporates study characteristics to model the variation among studies' effect sizes. Our model is a continuation of this line of thought and allows the inclusion of a scientific communities' beliefs about various methodologies' biases.
Tools for the formal assessment of randomized \citep{cochrane_ch8} and non-randomized \citep{cochrane_ch25} studies' risk of bias have been developed and are routinely used by practitioners. Our labeled random-effects model adds to the literature about incorporating prior knowledge about studies (such as information about their methodology) into meta-analysis to improve evidence synthesis~\citep{Li_94,Sutton_2001,Doi_Thalib_2008}.

We illustrate SMART by analyzing datasets drawn from two ends of the spectrum of methodological innovation: one literature is methodologically homogeneous, while the other experiences a noteworthy methodological advance.
SMART measures the relative influence of each study and helps illuminate the trajectory of scientific progress in the two cases. When the trajectory contains a methodological advance, SMART calls attention to scientific contributions that a conventional meta-analysis might overlook.

\section{Sequential Bayesian Meta Analysis for Tracking Scientific Consensus}
\label{Sec:QuantifyingLearning}

We develop a sequential meta-analysis method to track how evidence accumulates as new studies arrive. SMART sequentially updates a posterior distribution of the target parameter after each new study, and we quantify how much the posterior distribution changes at each step. Our method involves sequential updating with a Bayesian meta-analysis model, where each new study revises our collective beliefs, and a measurement of posterior change, which summarizes how those beliefs shift over time.

\subsection{Sequential Bayesian updating}
Consider a target parameter $\vartheta$, such as a causal effect that we are trying to estimate, and suppose that studies $y_t$ will appear sequentially. Sequential Bayesian updating revises our beliefs about $\vartheta$ as new studies appear.  Before observing any studies, we express initial uncertainty about $\vartheta$ with a prior distribution $p(\vartheta)$. After observing study $y_t$, we update the posterior distribution using Bayes' rule,
\begin{align}
  \label{eq:seq_update}
  p(\vartheta \mid y_{1:t})
  \propto p(y_t \mid \vartheta)\,p(\vartheta \mid y_{1:t-1}).
\end{align}
Here $p(y_t \mid \vartheta)$ is the likelihood model of a study, and $p(\vartheta \mid y_{1:t-1})$ is the posterior distribution after studies $1$ through $t-1$. Notice that the posterior after study $t$ becomes the prior before study $t+1$. These sequential updates formally accumulate scientific evidence across multiple studies.

Sequential updating requires a Bayesian meta-analysis model---a joint probability model $p(\vartheta, y_{1:t})$ that connects the parameter $\vartheta$ to the observed data. The posterior distribution, defined in Eq.~\ref{eq:seq_update}, is derived from this joint model. We next describe two hierarchical models for sequential Bayesian meta-analysis: the standard Bayesian random effects model (Section~\ref{sec:random-effects-model}) and an extension, which we call the labeled random effects model (Section~\ref{sec:labeled-random-effects-model}).

\subsection{Bayesian Random Effects for Meta Analysis}
\label{sec:random-effects-model}

The random effects model is a workhorse in the meta-analysis literature~\citep{dersimonian1986meta,Higgins_Thompson_Spiegelhalter_2009,Spence_Steinsaltz_Fanshawe_2016}. It describes studies whose measurements differ systematically from the target parameter $\vartheta$. Formally, study $t$ produces measurement $y_t$ according to the hierarchical model
\[
  y_t \sim \mathcal{N}(\vartheta + b_t, \sigma_t^2), \quad b_t \sim \mathcal{N}(0, \tau^2),
\]
where $b_t$ is the systematic bias of study $t$, $\sigma_t^2$ is the known sampling variance, and $\tau^2$ is the variance of biases across studies. Given a prior on the target parameter, $p(\vartheta)$, we sequentially update our posterior distribution as in Eq.~\ref{eq:seq_update}.

In the random effects model, the bias term $b_t$ can represent two related ideas. \textit{Statistical bias} is a systematic discrepancy between the expected value of an estimate and the true parameter. For example, observational studies can systematically differ from randomized experiments. \textit{Conceptual bias} is a systematic gap between what a study measures and the scientific target of interest. For example, studies on rats measure treatment effects that differ systematically from effects in humans; or studies from limited populations differ systematically from the full population of interest. Mathematically, conceptual and statistical biases are identical---each $b_t$ captures a systematic difference from $\vartheta$.

The random-effects model is appropriate for sequential meta-analysis because it allows our uncertainty about the target parameter to increase or decrease with each new study. A new study that aligns closely with previous findings typically reduces posterior uncertainty. But a surprising or conflicting study can increase posterior uncertainty, as it suggests larger systematic biases than previously believed. This flexibility distinguishes the random-effects model from the simpler fixed-effect model, which assumes no systematic biases and thus cannot increase uncertainty in response to conflicting evidence \citep{Cooper_2009}; see Appendix~\ref{sec:Meta_analysis}.

Though flexible, the random-effects model also makes strong assumptions. Specifically, through its prior distribution, it assumes that the biases
$b_t$ have mean zero. This assumption encodes the belief that studies are unbiased on average. But if some studies systematically differ, for example, due to known methodological biases, then this assumption is violated. (In classical meta-analysis, excluding studies that are known to be flawed is common practice.)

The zero-mean assumption might seem unpalatable, but note that the average bias of the studies is not identifiable from the data alone. Without additional assumptions, the parameter $\vartheta$ cannot be distinguished from the average bias across studies \citep{gustafson2015bayesian}. It is this identification problem that motivates the model of the next section.

\subsection{Labeled Random Effects for Meta Analysis}
\label{sec:labeled-random-effects-model}

Suppose we know specific study characteristics that correlate with biases, for example, study design, population sampled, or methodology. Then we can partially relax the zero-mean assumption by explicitly modeling these characteristics. The \textit{labeled random-effects model} allows systematic biases to depend on observable labels associated with each study. It provides additional structure that can help protect our estimates of the target parameter from systematic biases.

Formally, suppose each study $t$ is associated with a known methodological label $\ell_t$, indicating the methodology or category of the study, and suppose there are $L$ possible methodologies. Each methodology $k$ introduces a bias $\gamma_k$, which we model hierarchically as
\begin{align}
  \gamma_k &\sim \mathcal{N}(0, \kappa_k^2) \quad ( k = 1, \dots, L).
\end{align}
Here, $\kappa_k^2$ describes how much uncertainty we have about the bias associated with methodology $k$. A reliable methodology has a small $\kappa_k^2$, indicating little expected bias, whereas an uncertain or potentially biased methodology has a large $\kappa_k^2$.

With these labeled biases, our model of the observed measurement $y_t$ becomes
\[
  y_t \sim \mathcal{N}(\vartheta + \gamma_{\ell_t} + b_t, \sigma_t^2), \quad b_t \sim \mathcal{N}(0, \tau^2),
\]
where $b_t$ is the study-specific random effect as in the previous model, and $\gamma_{\ell_t}$ is the systematic bias due to the methodology of study $t$. Again, we can update the posterior sequentially with
Eq.~\ref{eq:seq_update}.
Beliefs about biases $\kappa^2_k$ may vary over time. When this is the case, we can quantify how much has been learned retrospectively (i.e., gauging the contribution of a study given our current understanding of biases). Alternatively, we can calculate the posterior at every time point with the $\kappa^2_k$'s describing the beliefs at the time. This gives us a picture of how influential studies were at the time they entered the literature.

This extension explicitly captures method-specific bias, and it allows us to incorporate expert knowledge about which methods are more or less trustworthy. Specifically, the degree to which we believe each methodology to be biased is encoded in $\kappa_{k}$. This, in turn, determines how much the studies that use that methodology contribute to the posterior of $\vartheta$.  The labeled random-effects model requires
that each study is labeled with its methodology, but it improves both
identification and interpretability of the parameter $\vartheta$.

In meta-analysis, studies that are deemed biased are often removed
\citep{Campbell_guidlines}. The labeled random-effects model provides a formal rationale for this often informal practice, down-weighting rather than excluding altogether data from methodologically questionable studies. Note also that it is
different from meta-regression
\citep{Thompson_Higgins_2002,Stanley_2005,stanley2012meta}, which aims
to explain the heterogeneity of reported effects, using study-level covariates.

Unlike the random-effects model, which assumes all studies to be exchangeable, the labeled random-effects model only assumes exchangeability within each label. Benefits of the relaxation of this assumption have been noted in \citep{Raudenbush_1985}.

Finally, we note that both the random-effects and labeled random-effects models rely on substantive prior assumptions. The random-effects model assumes the biases have zero mean; the labeled random-effects model further assumes that each methodology's bias has zero mean and that its variance ($\kappa_k^2$) is substantively informed. One might wonder whether these assumptions can be relaxed further, allowing data to determine the average bias or the variance parameters directly. Unfortunately, such an ``agnostic'' approach faces fundamental identification problems~\citep{gustafson2015bayesian,Gerber_Green_Kaplan_2004}: without prior constraints, the data cannot distinguish the target parameter $\vartheta$ from systematic shifts in bias or uncertainty about biases. Consequently, fully agnostic Bayesian meta-analysis is impossible. Priors---explicitly or implicitly---are unavoidable. The principled solution is sensitivity analysis, which transparently explores how conclusions about $\vartheta$ depend on prior assumptions~\citep{Li_Ding_Mealli_2022,Mikhaeil_2025}.

\subsection{Sequential Measurements of Posterior Change}
\label{Sec:LearningMetric}

Our goal is to sequentially update a collective posterior and measure how much our collective belief about the target parameter $\vartheta$ changes after each new study. Tracking posterior change reveals how evidence accumulation shapes our understanding of the target parameter over time. It can help identify influential studies, and quantify the point at which a scientific consensus begins to stabilize.

Formally, let $p_t \triangleq p(\vartheta \mid y_{1:t})$ denote the posterior distribution after observing studies $1$ through $t$. A natural way to quantify the distance between successive posteriors $p_t$ and $p_{t+1}$ is the Wasserstein-$p$ distance,
\begin{align}
\label{eq:LM}
W_p(p_t, p_{t+1}) = \left(\inf_{\gamma \in \Pi(p_t,\,p_{t+1})} \int |x_1 - x_2|^p \, d\gamma(x_1,x_2)\right)^{1/p}.
\end{align}
This metric can be interpreted as the minimal cost of transporting the mass from the previous posterior distribution $p_t$ to the updated posterior $p_{t+1}$~\citep{Villani_2008}.

The Wasserstein distance is particularly appealing for sequential meta-analysis. It is computationally convenient, especially when the posterior distributions are one-dimensional. Moreover, when the posteriors are normal, the Wasserstein-$2$ distance has a closed-form solution:
\begin{align}
W_2^2\bigl(\mathcal{N}(\mu_t,\sigma_t^2),\,\mathcal{N}(\mu_{t+1},\sigma_{t+1}^2)\bigr) = (\mu_{t+1} - \mu_t)^2 + (\sigma_{t+1} - \sigma_t)^2.\nonumber
\end{align}
In this expression, shifts in the posterior mean ($\mu$) and changes in the posterior standard deviation ($\sigma$) both contribute to the measure. The Wasserstein distance metric thus captures the possibility that an important study might not only refine our estimate but could also substantially change our uncertainty about the target parameter~\citep{mikhaeil2025b}.

\subsection{Tracking scientific consensus}

We have described an approach to sequential meta-analysis that involves three ingredients: (1) a Bayesian meta-analysis model, which formalizes assumptions about study biases and the target parameter; (2) a method for sequential Bayesian updating, derived from the model, which revises our collective beliefs about the target parameter after each new study; and
(3) a metric of posterior change, which quantifies how much each study contributes to scientific understanding. Together, these ingredients allow us to systematically analyze sequences of studies, explicitly track how scientific consensus emerges or shifts over time, and determine which studies were most important in forming that consensus. Next, we illustrate this framework using synthetic examples and real datasets.

\section{Studies}
\label{Sec:Studies}
We now illustrate our sequential Bayesian meta-analysis framework. Our
goal is to illustrate how the method tracks the evolution of
scientific evidence, quantifies the impact of new studies, and
explicitly models biases. Specifically, we address these
methodological questions: How do beliefs about a scientific parameter
evolve as studies accumulate? How does explicit modeling of systematic
biases influence our conclusions? How sensitive are results to our
prior assumptions?

We will begin with a large-scale replication dataset from psychology (Many Labs, Section \ref{Sec:ManyLabs}). SMART recovers the diminishing returns on research that are well-known for genuine (and successful) replications. 

We then move to a simulated example (Section \ref{Sec:ToyExample}) that illustrates the impact of methodological advances on a literature's trajectory. The advent of a new method may lead to a reevaluation of beliefs about past studies' biases.
 SMART picks up on the special role of pioneering studies and shows how their influence depends on a research community's beliefs (and their changes) about the biases of its methods.

We end with a case study in which we explore the importance of Card and Krueger's seminal 1994 study \citep{Card_Krueger_94}, which, due to its persuasive design, was essential in challenging prevailing economic thought on the effects of a minimum wage on employment.

\subsection{Many Labs}
\label{Sec:ManyLabs}
\begin{figure*}[t!]
    \includegraphics[width=\textwidth]{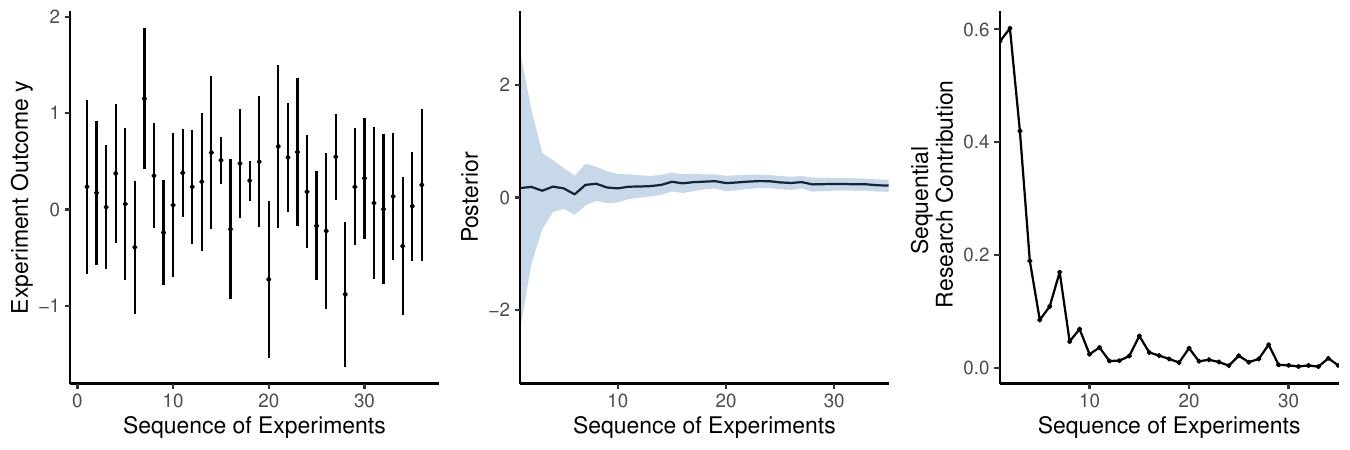}
   
   \caption{Sequential meta-analysis trace for the Many Labs example. (Left) Sequence of study outcomes replicating the ``Imagined Contact'' study \citep{Klein_2014}. (Middle) Evolution of the posterior mean with its 95\% credible interval. (Right) Sequential research contribution of every study, measured by the Wasserstein-2 learning metric. The trace recovers the diminishing returns that are to be expected when a single method is applied repeatedly.}
  \label{fig:ML_W2}
\end{figure*}
We begin by analyzing the Many Labs
dataset, a large-scale psychology replication project, which allows us
to track how evidence about certain psychological effects evolves as
studies from multiple laboratories accumulate. We analyze the original
Many Labs data, illustrating sequential posterior updating in
a real replication scenario. 
Data and code for all
examples may be found on \href{https://github.com/JonasMikhaeil/SequentialMATrace}{github}.

Each Many Labs study contains a collection of survey responses. The
design is described in \citep{Klein_2014}, and we focus on the
``Imagined Contact'' study. In the treatment condition, subjects are
asked to imagine interacting with a Muslim stranger for one
minute. (One exception is when the study is in Turkey, where they are
asked to imagine interacting with a Christian stranger.) In the control
condition, subjects are asked to imagine that they are walking
outdoors for the same length of time. After imagining their assigned
scene, subjects respond to a series of questions. The outcome variable
is a composite index of four questions that ask about interest in
learning about Islam and interacting with Muslims.

Specifically, each response is on a 1-9 scale, with higher values
indicating a more positive attitude, and the outcome is the average of
these four items. The studies were designed to be as similar as
possible in terms of the stimulus and outcome measures. So they
comport with the RE meta-analysis model, which attributes
study-to-study variation as a combination of statistical error and
idiosyncratic effect heterogeneity. Our goal is to estimate the
expected difference in outcome between treatment and control.

Figure \ref{fig:ML_W2} shows the sequential research trace for the
original sequence of studies. The initial studies contribute
substantially to our shared knowledge and significantly affect the
collective posterior. Eventually, however, the marginal contribution
of each study tapers off, and studies later in the sequence have
little effect on the posterior.

\subsection{A Simulation of Methodological Innovation}
\label{Sec:ToyExample}

Advances in research methodology can lead to surprising results. In
this section, we present a simulated example that illustrates how our
method tracks nonlinear scientific progress.  We simulate data in which
study outcomes $y_i$ are estimates of a parameter of interest
$\vartheta$, but contaminated by a method-specific bias $z_i$.

\begin{figure*}[t!]
    \includegraphics[width=\textwidth]{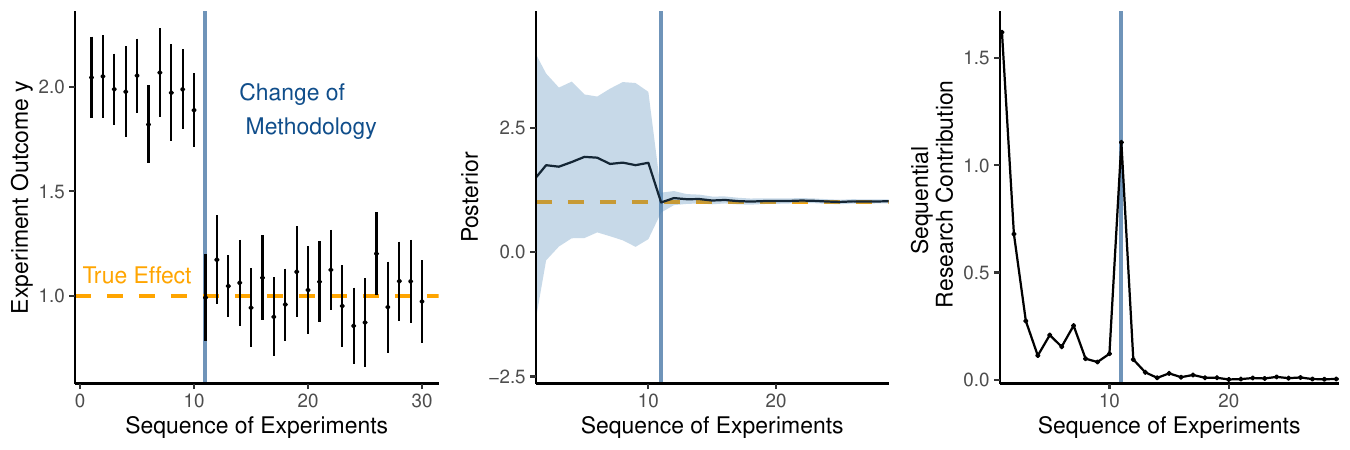}
  \caption{Sequential meta-analysis trace for the stylized example of methodological innovation. (Left) Sequence of experimental outcomes. The first 10 studies are biased. The 11th study pioneers improved and unbiased methodology that takes hold in the literature. (Middle) Evolution of the posterior mean and its uncertainty ($95$\% credible interval). Because the original methodology is believed to be biased, the posterior remains diffuse for the first 10 studies. By introducing an unbiased methodology, the 11th study shifts the posterior towards the true effect and significantly reduces posterior uncertainty. (Right) Sequential research contribution measured with the Wasserstein-2 learning metric. Due to its impact on the posterior both in terms of its location and scale, the 11th is highly influential. Our trace picks up on its significant research contribution.}
  \label{fig:syn_trace}
\end{figure*}

Specifically, we draw data from the following data-generating process
\begin{align}
   z_i \lvert \ell_i \, &\sim  \, \text{normal}(\beta \ell_i,.01) \\ \nonumber
   y_i \lvert z_i \, &\sim \, \text{normal}(\vartheta^* + z_i,.01).    
\end{align}
Here correlations in the research methodology of different labs
$\ell_i$ lead to correlations of the results $y_i$. We simulate a
scenario in which the first ten experiments ($1 \leq i \leq 10$) use a
specific and established methodology $\ell_i = 1$. After a while,
researchers develop a new and more accurate methodology $\ell_i = 0$,
and use it for the subsequent experiments ($11 \leq i \leq 30$). Figure \ref{fig:syn_trace} (Left) shows a sequence of 
experimental outcomes for this example.

Using the methodological labels, we analyze this data
with the labeled random effects model of
Section \ref{Sec:QuantifyingLearning}\ref{sec:random-effects-model}. What role do our beliefs
about the methods' accuracy (before and after the change of
methodology) play?

To demonstrate these methods, we consider two possible scientific
scenarios.  In the first scenario (I), we suspect the original
methodology to be biased (say, $\kappa_1 = 1$). Through conceptual
advances, the pioneering study (No. 11) introduces a new methodology
that the research community holds to be unbiased
($\kappa_2 = 0.0001 \ll \theta$). Figure \ref{fig:syn_trace}
shows the sequential meta-analysis trace for this example.
The middle panel tracks the evolution of the posterior. The right panel shows how much is learned by each study measured in terms of our
learning metric. The methodological shift to an unbiased method makes
the 11\textsuperscript{th} study highly influential.

\begin{figure*}[t!]
    \includegraphics[width=\textwidth]{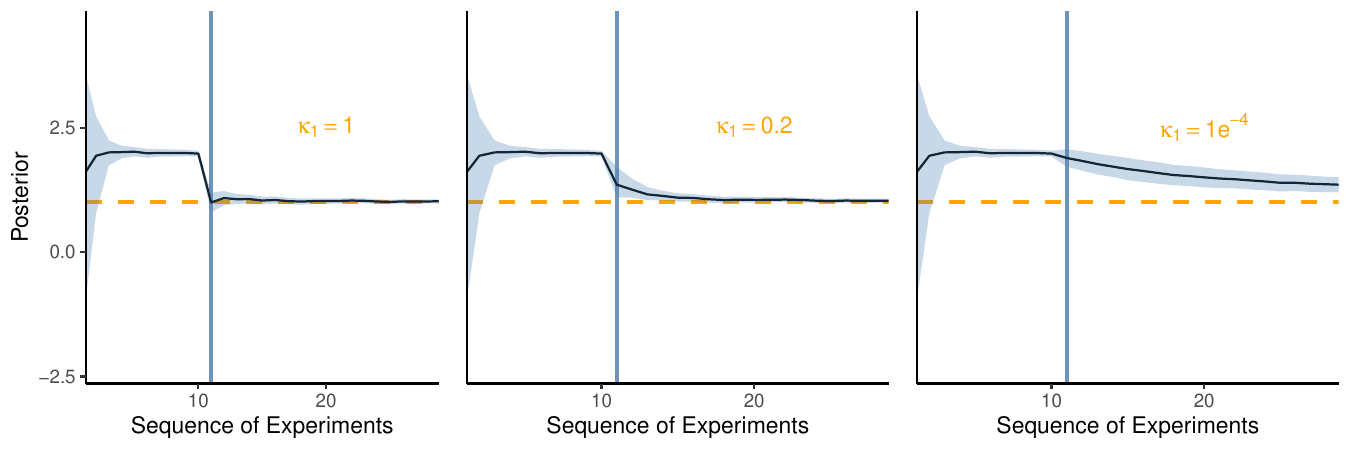}
    \caption{Evolution of the posterior for different beliefs about methodological biases. In all three situations, the methodology of the first 10 studies is originally believed to be unbiased ($\kappa_{i<11,1} =
0.0001$). The 11th study pioneers new unbiased methodology and advances theoretical reasons that cast doubt on the original methodology, changing the community's beliefs about the method ($\kappa_{i\geq 11,1} = \kappa_1$). The panels show the evolution of the posterior in the case that the beliefs change to (Left) $\kappa_1=1$, (Middle) $\kappa_1=0.2$ or (Right) remain unchanged $\kappa_1=0.0001$.}
  \label{fig:sens_syn}
\end{figure*}

In a second scenario (II), suppose that the first methodology is
originally believed to be accurate ($\kappa_{i<11,1} =
0.0001$). Theoretical and conceptual research, however, uncovers flaws
in this methodology, which eventually leads the research community to
acknowledge the possibility of bias for this methodology
($\kappa_{t\geq 11,1} = \kappa_1$). Building on this development, a
pioneering study (No. 11) introduces a new method into the literature,
and this method is believed to yield unbiased results
($\kappa_{i\geq 11,2} = 0.0001$).  Figure \ref{fig:sens_syn} tracks the posterior for different degrees $\kappa_1$ to which the research community
begins to doubt the first methodology. The middle and right panel show that posterior uncertainty can
increase when a new research methodology establishes itself even if
beliefs about methodological biases are not taken into
account. Studies that pioneer a change in methodology can be
pathbreaking, because they cast doubt on what has previously been
learned. The figure also highlights how much is learned from the pioneering study (No. 11)
depends on the degree $\kappa_1$ to which the research community
begins to doubt the first methodology.

Both scenarios highlight the importance of innovation in
research. Through the development of more accurate methods, there can
be a high return to a study in terms of learning even when the same
literature would face diminishing returns sticking to a single
method. Our second scenario highlights the contribution of theoretical
work (that may not contribute any data at all) to the importance of
pioneering empirical studies by informing our beliefs about a
literature's potential biases

Finally, we use this simulation to illustrate the difference between
our sequential meta-analysis trace and classical meta-analysis
weights \citep{Riley_Ensor_Jackson_Burke_2018,Burke_2018}.\footnote{For a definition of meta-analysis weights, and a more in-depth discussion, see Appendix \ref{Sec:CompMAW_SMAT}.} Table \ref{tab:MAweights} (in Appendix \ref{Sec:CompMAW_SMAT}) contains
meta-analysis weights calculated sequentially and retrospectively for our stylized example. Despite pioneering novel methodology, study 11 receives no outstanding weight.
Through the lens of meta-analysis, the value of a study reflects its contribution to the overall meta-analysis effect estimate. 
No account is taken of whether certain studies may have been ground-breaking by advancing new methodologies. 
When meta-analysis weights are used to gauge a study's contribution to the literature, studies with small reported standard errors receive higher weights. This essentially favors large (and oftentimes expensive) studies while diminishing the importance of smaller and innovative pilot studies. Through the lens of our method for quantifying learning, the precision of a study is only one factor contributing to its impact. By being able to quantify the value of innovation, our method gives due credit to pioneers. 

\subsection{Case Study: Assessing the Employment Effects of Minimum Wage Laws}
\label{Sec:CaseStudy}

\begin{figure*}[t!]
    \includegraphics[width=\textwidth]{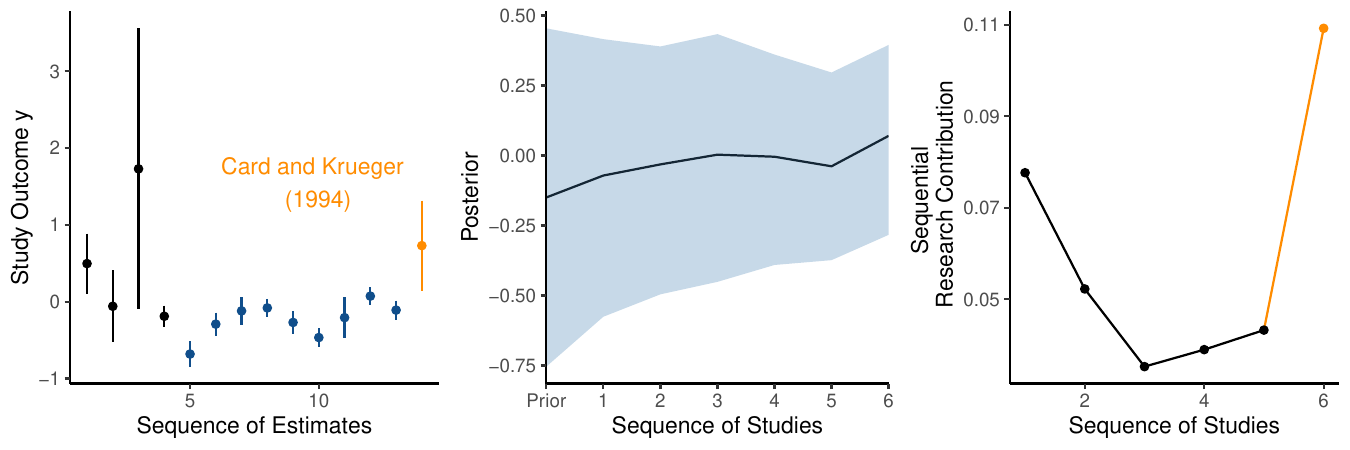}
    \caption{Sequential meta-analysis trace for the minimum wage literature leading up to Card and Krueger's 1994 study \citep{Card_Krueger_94}. 
    We begin with a prior $\pi_0(\theta) = \mathcal{N}(-0.15,0.07^2+0.3^2)$ based on discounted time-series evidence \citep{Brown_82}. We assume that the methodologies used before C\&K's study were believed to be biased ($\kappa_{1} = \kappa_{2} = \kappa_{3} = 0.3$) and that C\&K's border discontinuity design was believed to be less biased ($\kappa_4 = 0.05)$. 
    (Left) Sequence of estimates leading up to C\&K's study (with their estimate in
      orange). Estimates colored in blue belong to the same study \citep{Williams_93} and the posterior is updated based on them collectively. (Middle) Evolution of the posterior mean and its $95$\% credible interval. (Right) Sequential research contribution measured with the Wasserstein-$2$ learning metric. This trace recovers the influential role C\&K's study played in the New Minimum Wage Research \citep{Ehrenberg_92} literature.  }
  \label{fig:MW_figure}
\end{figure*}
\begin{figure}[th]
\centering
\includegraphics[width=0.7\textwidth]{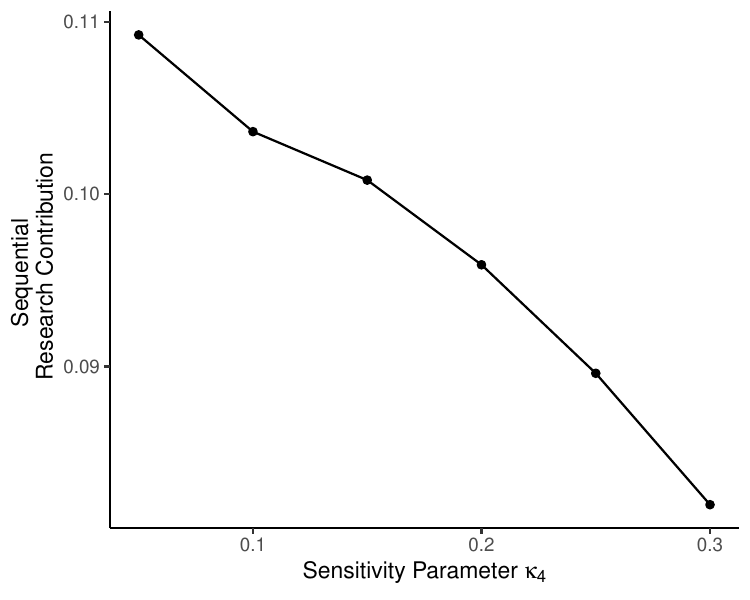}
\caption{Sensitivity analysis for C\&K's 1994 study's \citep{Card_Krueger_94} sequential research contribution. Their contribution crucially depends on how unconfounded their border discontinuity design is believed to be. For values of $\kappa_4$ close to $0$ C\&K's results is credible. With increases in credibility, a more significant sequential research contribution is accorded to C\&K's study.}
\label{fig:MW_sensitivity}
\end{figure}
In this section, we present a real-world illustration of our method to
assess sequential research contributions. We focus on a literature estimating the effect of a minimum wage on employment and, specifically, seek to understand the role of the seminal 1994 study by Card
\& Krueger \citep{Card_Krueger_94} --- we will refer to them as C\&K --- who
used the 1992 increase in New Jersey's minimum wage to estimate
this effect. 
The minimum wage is an important policy lever intended to increase income among low-wage
earners. Economists worry, however, that raising the minimum wage may negatively affect
employment, hurting some low-wage workers while helping others. 
The best evidence prior to the 1990s, which was based on time-series data, suggested that increases in the minimum wage led to a reduction in employment, a result in line with the textbook microeconomics.
Labor economists by the late 1980s, however, increasingly discounted these empirical results because of potential confounding.
The 1990s saw the ascent of a \textit{New Minimum Wage Research} literature
\citep{Ehrenberg_92}, featuring innovative study designs that would challenge classical economic thought on this topic \citep{myth_95}. 

We look at the beginning of this literature, which featured a methodological shift from
time-series analysis to estimation strategies leveraging plausibly exogenous variation in the minimum wage. 
The goal of the new
approach was to reduce the threat of omitted variables bias and thus provide more credible estimates. Specifically, we analyze the trajectory of six
studies from 1992 to 1994 leading up to, and including, C\&K as presented by Neumark and Shirley
\citep{Neumark_Shirley_22}.\footnote{
All estimates report elasticities. For studies reporting multiple estimates of the same effect with different controls \citep{Cardb_1992,Neumark_1992}, we selected one representative estimate.} Figure \ref{fig:MW_figure} shows the sequential research trace for this literature. 

One of the early contributions to this new research literature was C\&K's
classic study \citep{Card_Krueger_94}, which rose to prominence because of its convincing design. To mitigate confounding, the authors applied 
difference-in-differences (DID) estimation to employment data gathered along a border discontinuity, comparing counties in New Jersey affected by the increase of the state's minimum wage with adjacent unaffected counties in Pennsylvania  \citep{nobel_21}. Another compelling aspect of this design was its prospective collection of data, which reduced researcher degrees of freedom.\footnote{Card was awarded
  the Nobel Memorial Prize in Economic Sciences partly for this and
  related work. DID itself was introduced to the minimum
  wage literature in 1915 by Obenauer and von
  der Nienburg \citep{Obenauer_Nienburg}, see
  \citep{Angrist_Pischke}. Card also used DID to estimate the effect of a minimum wage increase in California \citep{Card_92}.}

There is widespread agreement that C\&K's study was highly influential.\footnote{To give just one quotation
  highlighting C\&K's study perceived influence: ``The path-breaking
  work of Card and Krueger [1994], showing that a higher minimum wage
  can increase employment, turned the age-old conventional wisdom on
  its head.''  \citep{marjit_21}} Their study has become a widely-cited
textbook example of a well-designed study with enormous policy impact \citep{Angrist_Pischke}. It also inspires ongoing
theoretical \citep{Dube_16,Card_18,marjit_21,Renkin_22} and methodological advances
\citep{Neumark_13,Abadie_2005,Athey_2006,torous2024optimaltransportapproachestimating}.

Classical meta-analysis, however, accords it little weight. Applying the random-effects model to the sequence of studies from 1992 to 1994, we find that C\&K's study is given a weight of 15\%, which is the second smallest weight in this collection of studies. Unlike classical meta-analysis, the SMART analysis in Figure \ref{fig:MW_figure} picks up on the seminal impact of C\&K.

Our aim is not to weigh in on the employment effects of the minimum
wage, which continue to be the focus of research and
debate \citep{Neumark_Shirley_22,Manning_21}. Rather, we use our
sequential meta-analysis trace to understand the influence of C\&K's study on the development of
this literature. 

What made C\&K's study so influential?  First, C\&K's study found a surprising result. This alone,
however, would not have been enough. It is only in conjunction with
the research community's widespread belief that
C\&K's methodology is more accurate than previous methods that
their surprising results shifted the collective posterior.

To formulate a research trace for this literature, we begin by reconstructing a prior. The time-series evidence up to the early 1980s is summarized in \citep{Brown_82}, which concludes that negative elasticities between 1$\%$ and 3$\%$ are most plausible with the lower range being more likely. A rough approximation of this summary is a prior such as $\mathcal{N}(-0.15,0.07^2)$.  
A growing concern with confounding (see, for example, \citep{LaLonde_86})  led labor economists in the late 1980s to discount this time-series evidence. Somewhat arbitrarily, we can widen the prior $\pi_0(\theta) = \mathcal{N}(-0.15,0.07^2+0.3^2)$ to express this additional uncertainty, but note that the story our trace conveys does not depend on the exact illustrative numbers chosen.

To quantify the influence of C\&K's study, we again use the labeled
random-effects model of Section \ref{Sec:QuantifyingLearning}\ref{sec:random-effects-model}.  We group all methodologies using panel data on federal variation of the minimum wage under $\ell=\ell_{fed}$ \citep{Cardb_1992,Katz_92,Williams_93} and those using state-level variation under $\ell=\ell_{st}$ \citep{Neumark_1992}. $\ell=\ell_{DID}$ denotes the DID methodology used in \citep{Card_92} and $\ell =\ell_{CK}$ indicates C\&K's border discontinuity design \citep{Card_Krueger_94}.

While the studies based on (observational) panel data were all aiming to leverage plausibly exogenous variation in the minimum wage, the threat of omitted variables bias remained a lingering concern \citep{nobel_21}.
For simplicity, we assume that all studies predating C\&K were believed to be biased to the same degree $\kappa_{1} = \kappa_{2} = \kappa_{3} = 0.3$. To determine the effect of the minimum wage on employment, the effects of the change in the minimum wage need to be disentangled from other factors affecting employment. 
By comparing adjacent counties that experience similar economic shocks, some affected by a rise in the minimum wage while others remained unaffected, C\&K's border discontinuity design 
plausibly mitigated confounding. This innovative design led to the belief that this study was less biased than its predecessors, say, $\kappa_4= 0.05$.

Figure \ref{fig:MW_figure} showcases the significance of C\&K's study. 
On the left, we see the sequence of estimates leading up to C\&K's study. The middle panel tracks how the posterior is sequentially updated with every study. The plot on the right shows each study's sequential research contribution as it enters the literature, with C\&K's study clearly standing out for its importance.

Figure \ref{fig:MW_sensitivity} provides a sensitivity analysis for assessing the importance of C\&K's study
\citep{Card_Krueger_94}. It shows how C\&K's study's research contribution decreases as a function of $\kappa_{4}$: the more convincingly their design overcomes confounding, the more influential their study becomes. If \citep{Card_Krueger_94} were just another flawed study, its influence would have been unremarkable.

\section{Conclusion}

We propose the sequential meta-analysis research trace (SMART). We take a sequence of studies, each one targeting the same parameter, and containing an estimate and a standard error. Applying a Bayesian approach, we plot sequentially updated 
posterior estimates of the target parameter and calculate each study's research contribution,
defined as the change in
its posterior relative to the prior literature. 
Our approach complements classical methods for meta-analysis.
Classical meta-analysis analyzes all the studies simultaneously to
produce a blended estimate of the target parameter, as well as weights that describe each study's
retrospective importance. Here we analyze the studies sequentially,
which leads to the same blended final estimate but assesses the importance
of each study at the time it appeared.

We are by no means the first to consider or conduct sequential meta-analysis; other work also updates a Bayesian meta-analysis model one study at a time \citep{Higgins_Whitehead_Simmonds_2011,Van_Der_Tweel_Bollen_2010,Bollen_Uiterwaal_Van_Vught_Van_Der_Tweel_2006}. But we additionally quantify the change in posterior, proposing that meta-analysts use the Wasserstein distance for gauging the importance of a study at the time it appeared.  This yardstick for quantifying research contributions can recognize the substantial impact of small studies that move a literature in a distinctive direction --- studies that would ordinarily be accorded little weight by a random effects meta-analysis, whether sequential or retrospective.  Although our approach to meta-analysis does not presuppose a specific meta-analytic model, we believe that models that take note of methodological ``labels’’ may offer insights that might otherwise be missed when summarizing literatures with active methodological debates.

Our simulations and empirical example show that methodological innovations are characterized very differently by models that explicitly acknowledge the possibility that new insights may impeach the evidence that came before.
The profound influence of the Card and Krueger minimum wage study, for example, reflects its methodological insights rather than the precision of its empirical estimate.  In the social sciences, methodological upheaval is not simply an abstract possibility; one cannot appreciate the development of influential research literatures without taking notice of the sharp break that occurred due to the advent of natural experiments \citep{Ehrenberg_92}, regression discontinuity designs \citep{Thistlethwaite_60,Imbens_08}, or randomized control trials \citep{Angrist_10,Banerjee_16}.

We emphasize that SMART is a way of looking at a research literature.
Like all model-based methods, it comes with assumptions and
simplifications. We assume the parameter of interest is fixed over
time and that the sequence of studies we encounter constitute independent random draws from a sampling distribution.  We make the same parametric assumptions about study
likelihoods and priors that are posited in the classical
random-effects meta-analysis model, but when applying a ``labeled’’ meta-analysis model, we further assume that we can quantify the relative uncertainty over a set of 
methodological biases. 

That said, the model that underlies SMART provides potential insights
into the trajectory of a scientific literature. It can detect when a
literature loses momentum as it converges on a consensus answer with
increasing precision. Or it can show that a literature has been
revitalized by methodological innovations that expose the deficiencies
of prior approaches. With an extension to a dynamic model, SMART could further detect a changing scientific landscape--where the underlying estimand changes, be it due to temporal effects or a change in the scientific question itself.

\bibliographystyle{plainnat}
\bibliography{literature}
\section*{Acknowledgments}
The authors are grateful to Columbia University for research support. All authors contributed to writing and revision.  DB and DG framed the central arguments, and JM took the lead in working out the technical exposition. 
The authors declare that they have no competing interests. All data needed to evaluate the conclusions in the paper are present in the paper. Code to reproduce our results is available on \href{https://github.com/JonasMikhaeil/SequentialMATrace}{https://github.com/JonasMikhaeil/SequentialMATrace}. The authors acknowledge that they received no funding in support for this research.
\newpage
\appendix
\section*{Appendix}
\section{Sequential Meta-Analysis}
\label{sec:Meta_analysis}
Here we want to briefly give a summary of the fixed-effect and random-effects models used in (sequential) meta-analysis. 
Classical \citep{Whitehead_2002,Cooper_2009} and cumulative \citep{Higgins_Whitehead_Simmonds_2011} meta-analysis are concerned with the 
aggregation of (unbiased) effect estimates. In this setup, every study reports an effect estimate $\hat \vartheta_i$ and corresponding standard error estimate $\hat \sigma_i$.

Fixed-effect meta-analysis\footnote{One can distinguish between the common effect and fixed-effects (plural) analysis in which the study effects are unknown and fixed but need not necessarily be identical. While the interpretation of $\vartheta$ changes, the estimators remain identical \citep{Rice_2018}.} builds on the assumption that the effect is constant across studies. In this case, the following estimates and standard error for the aggregate effect estimate 
\begin{align}
\label{eq:FE}
    \hat \vartheta_{\text{FE}} = \frac{\sum_{i=1}^n \hat \vartheta_i/\hat \sigma_i^2}{\sum_{i=1}^n 1/\hat \sigma_i^2}
\hspace{1cm} \text{and}\hspace{1cm} 
   \widehat{\text{Var}} (\hat \vartheta_{\text{FE}}) = \frac{1}{\sum_{i=1}^n 1/\hat \sigma_i^2}
\end{align}
are widely used. This estimator can be understood as weighted-least squares (with the inverse variance as weights) taking the variance estimates of the individual studies at face value (i.e., assuming that the variance of the variance estimate is negligible).
Alternatively, and in line with the Bayesian perspective taken in this paper, we can understand the fixed-effect model as 
\begin{align}\nonumber
    \hat \vartheta_i \lvert \vartheta \sim \text{normal}(\vartheta, \sigma^2_i).
\end{align}
The posterior of the fixed effect (assuming conditional independence) is given by
\begin{align}\nonumber
    p(\vartheta \lvert \hat \vartheta_1, \dots , \hat \vartheta_n) = \frac{\left(\prod_{i=1}^n \text{normal}(\vartheta, \sigma^2_i)\right) \pi(\vartheta)}{p(\hat \vartheta_1, \dots , \hat \vartheta_n)}.
\end{align}
If the prior $\pi(\vartheta)$ is chosen to be flat, the posterior mean $\mathbb{E}[\vartheta \lvert \hat \vartheta_1, \dots , \hat \vartheta_n] = \hat \vartheta_{FE}$ is equivalent to the classical fixed-effects estimator in Equation \ref{eq:FE} (again assuming the variance of the variance estimates to be negligible, i.e. assuming $\sigma_i$ to be known).

While effect homogeneity may be a reasonable assumption in some settings, it is doubtful in biomedical settings or the social sciences. Studies vary in numerous ways including their target population, their intervention, and their outcome measures. For these reasons, at least some degree of effect heterogeneity is to be expected \citep{Higgins_Thompson_Spiegelhalter_2009}. 
Following this line of thought, random-effects models recommend themselves.\footnote{There are various interpretations of random effects models \citep{Higgins_Thompson_Spiegelhalter_2009,Cooper_2009,Gelman_2005}. While we prefer the Bayesian interpretation below, random effects are also frequently understood by reference to a sampling mechanism by which effects for every study are sampled from a population of possible effects.} 
If we assume that the effects $\vartheta_i$ of the different studies are exchangeable - that is, they may be non-identical but we have no reason to distinguish their magnitude a priori - then de Finetti's theorem allows us to think of the $\vartheta_i$ as identically and independently distributed conditional on some (prior) parameter $\psi$ \citep{Bernardo_1994}.\footnote{More specifically, the sequence ${\vartheta_i}$ needs not only to be exchangeable but embeddable in an infinite sequence that is exchangeable.} 
This motivates the random-effects model \citep{Whitehead_2002}.

\begin{align}\nonumber
    \hat \vartheta_i \lvert \mu_i\, & \sim \,\text{normal}(\mu_i , \sigma^2_i) \\\nonumber
    \mu_i \lvert \vartheta,\tau \, & \sim \,\text{normal}(\vartheta,\tau^2), 
\end{align}
where $\vartheta$ is the mean of the study effects and $\tau$ describes the level of effect heterogeneity.
Integrating out $\mu_i$, we have 
\begin{align}\nonumber
    \hat \vartheta_i \lvert \vartheta, \tau \, & \sim \,\text{normal}(\vartheta, \sigma_i^2+\tau^2). 
\end{align}
The posterior of the mean of the study effects is then given by (assuming $\vartheta$ is independent of $\tau$)
\begin{align}\nonumber
     p(\vartheta \lvert \hat \vartheta_1, \dots , \hat \vartheta_n,\tau) = \frac{\left(\prod_{i=1}^n \text{normal}(\vartheta, \sigma_i^2+\tau^2)\right) \pi(\vartheta)}{p(\hat \vartheta_1, \dots , \hat \vartheta_n,\tau)}.
\end{align}
Choosing a flat prior for $\vartheta$ (assuming an appropriate plug-in estimator $\hat \tau$ for $\tau$ is available and neglecting the variance of $\hat \sigma_i$) yields the typically used random-effects estimator \citep{Whitehead_Whitehead_1991,Higgins_Thompson_Spiegelhalter_2009}:
\begin{align}
\label{eq:REest}
    \hat \vartheta_{RE} = \frac{\sum_{i=1}^n\frac{1}{\hat\sigma_i^2 + \hat \tau^2 }\hat \vartheta_i}{\sum_{i=1}^n\frac{1}{\hat\sigma_i^2 + \hat \tau^2 }} \hspace{0.3cm} \text{and}\hspace{0.3cm} 
   \widehat{\text{Var}} (\hat \vartheta_{RE}) = \frac{1}{\sum_{i=1}^n\frac{1}{\hat\sigma_i^2 + \hat \tau^2 }}.
\end{align}

From the discussion above, it is clear what the fixed- and random-effects estimates are when these models are applied sequentially. In this case we understand $\{(\hat \vartheta_i,\hat \sigma_i)\}_{i=1}^n$ as a sequence that we observe up to point  $m \leq n$. To perform meta-analysis sequentially, we apply the above models to $\{(\hat \vartheta_i,\hat \sigma_i)\}_{i=1}^m$ for every $1\leq m \leq n$.\footnote{As discussed in Section \ref{Sec:QuantifyingLearning}\ref{sec:random-effects-model}, the posterior of the random-effects model can also be updated in fully Bayesian fashion (that is, without the use of a plug-in estimator) using modern probabilistic programming languages, such as \citep{stan}}.

This  reveals that the fixed-effects model is
not conceptually suited to explain the kind of increased uncertainty observed in Section \ref{Sec:Studies}\ref{Sec:ToyExample}. As can be seen from Equation \ref{eq:FE}, when performing fixed-effects meta-analysis, every future study will increase our certainty (regardless of how surprising its outcome may be). 

For random-effect models, uncertainty may increase if a future study increases our estimate of effect heterogeneity $\hat \tau$  (see Equation \ref{eq:REest}). Classically, this increase in uncertainty, however, is solely attributed to effect heterogeneity. In Section \ref{Sec:QuantifyingLearning}\ref{sec:random-effects-model}, we offer an alternative interpretation of the random-effects model in terms of methodological biases. This allows us to understand the increase in posterior variance when the posterior is updated sequentially as a consequence of a new study forcing us to question our beliefs about the corpus of studies collected so far.  

The example in Section \ref{Sec:Studies}\ref{Sec:ToyExample} illustrates one case in which uncertainty may increase through a process other than effect heterogeneity.  In this example, revelations about systematic measurement error lead to a sudden change in posteriors.
While measurement error has been discussed in the meta-analysis literature, 
it is usually assumed to only attenuate effect estimates \citep{Cooper_2009}. Recent work, however, highlights that correlated measurement error can systematically bias effect estimates \citep{Knox_Lucas_Cho_2022,Mikhaeil_2025}. The wider issue of bias has been addressed in the meta-analysis literature \citep{Higgins_Thompson_Spiegelhalter_2009,Carter_2019,Wiernik_Dahlke_2020}, and both careful assessment of bias and selection of studies to be included are common.

\section{Comparison of Meta-Analysis Weights with our Sequential Meta-Analysis Trace}
\label{Sec:CompMAW_SMAT}
\begin{table}\renewcommand\thetable{S1} 
\centering
\begin{tabular}[t]{c|cc|cc}
Number in  & Sequential & Sequential  & Retrospective & Retrospective\\
Sequence & FE weight & RE weight & FE weight & RE weight\\
\midrule
1 & 100.0 & 100.0 & 3.3 & 3.3\\
2 & 49.4 & 49.4 & 3.2 & 3.3\\
3 & 39.7 & 39.7 & 4.3 & 3.4\\
4 & 19.8 & 19.8 & 2.7 & 3.3\\
5 & 22.9 & 22.9 & 4.0 & 3.4\\
6 & 17.1 & 17.1 & 3.6 & 3.3\\
7 & 11.6 & 11.6 & 2.8 & 3.3\\
8 & 8.8 & 8.8 & 2.3 & 3.3\\
9 & 11.6 & 11.6 & 3.4 & 3.3\\
10 & 11.8 & 11.8 & 4.0 & 3.4\\
11 & 7.9 & 9.0 & 2.9 & 3.3\\
12 & 7.1 & 8.2 & 2.8 & 3.3\\
13 & 12.6 & 7.9 & 5.7 & 3.4\\
14 & 6.2 & 7.1 & 3.0 & 3.3\\
15 & 6.9 & 6.7 & 3.5 & 3.3\\
16 & 5.6 & 6.2 & 3.1 & 3.3\\
17 & 6.1 & 5.9 & 3.5 & 3.3\\
18 & 6.8 & 5.6 & 4.2 & 3.4\\
19 & 4.1 & 5.2 & 2.6 & 3.3\\
20 & 4.2 & 5.0 & 2.9 & 3.3\\
21 & 4.7 & 4.8 & 3.3 & 3.3\\
22 & 4.6 & 4.6 & 3.4 & 3.3\\
23 & 4.3 & 4.4 & 3.3 & 3.3\\
24 & 4.6 & 4.2 & 3.7 & 3.4\\
25 & 3.3 & 4.0 & 2.8 & 3.3\\
26 & 3.7 & 3.8 & 3.2 & 3.3\\
27 & 2.9 & 3.7 & 2.6 & 3.3\\
28 & 3.7 & 3.6 & 3.5 & 3.3\\
29 & 3.3 & 3.4 & 3.2 & 3.3\\
30 & 3.2 & 3.3 & 3.2 & 3.3\\
\end{tabular}
\caption{\label{tab:MAweights} Fixed-effect (FE) and random-effects (RE) meta-analysis weights for the stylized example in Section \ref{Sec:Studies}\ref{Sec:ToyExample}. For the sequential meta-analysis weights, meta-analysis is performed sequentially and the weight of the most recent study is reported. The retrospective weights are obtained by performing meta-analysis on the entire sequence of studies. Meta-analysis weights, either computed sequentially or retrospectively, are not able to capture the influential nature of 11th study. }
\end{table}

Our sequential meta-analysis method to quantify how much has been learned is complementary to the classical meta-analysis perspective. Through the lens of meta-analysis, the value of a study reflects its contribution to the overall meta-analysis effect estimate. This view judges a study in comparison to all other studies as a function of each study's estimated standard error and the overall degree of effect heterogeneity. No account is taken of whether certain studies may have been ground-breaking by advancing new methodologies. Ironically, from the vantage point of (retrospective) meta-analysis, the fact that an innovative study inspired other studies to follow suit means that an innovative study will in hindsight obscure its own central role in the exploration of a phenomenon.  

To make this more formal, let us revisit the stylized example from Section \ref{Sec:Studies}\ref{Sec:ToyExample}. 
The introduction of new research methodology by the $11$\textsuperscript{th} study led to a drastic shift and widening of the collective posterior. In our hypothetical sequence of studies, this innovation also led future researchers to adopt the new methodology. This study was clearly highly influential. What do meta-analytic algorithms say about this study's importance?  

When performing meta-analysis, it is typical to look at the weights of studies (i.e., how much each study contributed to the pooled meta-analytic estimate) to determine their importance. Their weights are given by 
\begin{align}\nonumber
    \delta_i = \frac{w_i}{\sum_{i=1}^n w_i},
\end{align}
where $w_i = \frac{1}{\hat \sigma_i^2}$ for fixed-effect or $w_i = \frac{1}{\hat \sigma_i^2 +\hat \tau^2}$ for random effects meta-analysis. Here   $\hat \sigma_i$ are the standard errors reported by the studies and $\hat \tau$ is the meta-analysis estimate for effect heterogeneity \citep{Higgins_Thompson_2002}. 
To isolate the effect of innovation, our stylized example is set up in such a way that all studies' effects have the same standard error. We can thus immediately see that all studies will be accorded same weight -- even though study 11 was a pathbreaking and influential study, it appears as no different than any other study. Table \ref{tab:MAweights} shows the weights of the latest study when meta-analysis is performed sequentially. In stark contrast to our learning metric (see Figure  \ref{fig:syn_trace} (Right)), the meta-analysis weights imply diminishing returns.

Meta-analysis weights, even when meta-analysis is performed sequentially, fail to properly quantify the contribution of a methodologically innovative study. Innovation, as opposed to pure methodological diversity, is inherently sequential and the value of innovation cannot be captured purely by the corresponding studies' contribution to the aggregate effect estimate. By focusing on the effect a study has on the collective posterior at each point in time, which may drastically widen when innovation leads to a surprising outcome, our method of quantifying learning is able to pick up on the special role the $11$\textsuperscript{th} study plays in the stylized sequence of studies. 

While our discussion has centered on a stylized example, it has important implications for judging the value of studies in the real world. When meta-analysis weights are used to gauge a study's contribution to the literature, studies with small reported standard errors receive higher weights. This essentially favors large (and oftentimes expensive) studies while diminishing the importance of smaller and innovative pilot studies. Through the lens of our method for quantifying learning, the precision of a study is only one factor contributing to its impact. By being able to quantify the value of innovation, our method gives due credit to pioneers.

\section{Lindley's Learning Metric}
\label{Sec:Lindley}
In his seminal paper, Lindley \citep{Lindley_1956} develops a measure of how much information a study has provided: 
\begin{align}
    \mathcal{I}(\pi(\vartheta),y) \, &= \, H(\pi(\vartheta\lvert y)) - H(\pi(\vartheta)) \\ \nonumber
    \, &:= \,\mathbb{E}_{\pi(\vartheta\lvert y)}\bigg [ \log\pi(\vartheta\lvert y) \bigg] - \mathbb{E}_{\pi(\vartheta)}\bigg [ \log\pi(\vartheta) \bigg]
\end{align}
This expression depends on the observed data $y$; some experimental outcomes are more informative than others. In addition to its information-theoretic appeal, Lindley's measure $I(\pi(\vartheta),y)$ can also be understood as the reduction of risk corresponding to a decision problem where one has to decide which distribution to report about an unknown quantity $\vartheta$ \citep{Parmigiani_1994}. The utility function $U(\vartheta, \varphi) = \log \varphi(\vartheta)$ implied by Lindley's measure is appealing from this decision-theoretic perspective as it induces decision makers to report their current beliefs in the form of their posterior.

While theoretically appealing, Lindley's measure $I(\pi(\vartheta),y)$ can be hard to compute in practice \citep{Batu_Dasgupta_Kumar_Rubinfeld_2002}. 
Similar to the Wasserstein-2 distance, see Section \ref{Sec:LearningMetric}, Lindley's measure of information has a closed-form solution in the situation in which prior and posterior are normal. Let $\pi(\vartheta\lvert y) = \text{normal}(\mu_{post},\sigma_{post})$ and $\pi(\vartheta) = \text{normal}(\mu_{prior},\sigma_{prior})$, then 
\begin{align}
    \mathcal{I}(\pi(\vartheta),y) = \log{\sigma_{post}}-\log{\sigma_{prior}}.
\end{align}
This reveals why Lindley's information measure is unappealing in practice. Realistically, the sequential meta-analysis model does not coincide with the true data-generating process. In this situation, we are not guaranteed that the posterior mean concentrates on the true value of $\vartheta$ -- and different studies (for example by removing bias) may impact our shared knowledge by drastically changing the posterior mean. We thus recommend using the Wasserstein distances as learning metrics (Equation \ref{eq:LM}) as they are sensitive to changes in the entire posterior distribution.

\end{document}